\documentclass[aps,pre,preprint,twoside,floatfix,superscriptaddress,showpacs]{revtex4}
\usepackage{amsmath}
\usepackage{amsfonts}
\usepackage{amssymb}
\usepackage{graphicx}
\usepackage{overpic}
\usepackage{psfrag}
\usepackage{pifont}
\usepackage{color}

\begin{document}
%
\title{Effective interactions and equilibrium configurations of colloidal particles on a sessile droplet}
\author{J.~Guzowski}
\affiliation{Institute of Physical Chemistry, Polish Academy of Sciences, ul.\ Kasprzaka 44/52, 01--224 Warszawa }
\author{M.~Tasinkevych}
\author{S.~Dietrich}
\affiliation{Max--Planck--Institut f\"ur Metallforschung,
Heisenberstr.\ 3, 70569 Stuttgart, Germany}
\affiliation{Institut f\"{u}r Theoretische und Angewandte Physik,
         Universit\"{a}t Stuttgart, Pfaffenwaldring 57,
         D-70569 Stuttgart, Germany}

\date{\today}

\begin{abstract}
We study the free energy landscapes of a pair of submicron spherical particles floating at the surface of a sessile droplet. The particles are subjected to radial external forces resulting in a deformation of the droplet shape relative to the reference shape of a spherical cap. This deformation leads to tangential forces on the particles. For small deformations and for the contact angle $\theta_0$ at the substrate being equal to $\pi/2$, the corresponding linearized Young-Laplace equation is solved analytically. The solution is constructed by employing the method of images from electrostatics, where each of the particles plays the role of a capillary monopole and the substrate is replaced by a virtual drop with image charges and by imposing the conditions of fixed droplet volume and vanishing total force on the droplet. The substrate boundary conditions determine the signs of the image capillary charges and therefore also the strength of the tangential forces on the particles. In the cases of an arbitrary contact angle $\theta_0$ these forces are calculated numerically by employing a finite element method to find the equilibrium shape of the droplet for those configurations in which the particles are close to the local free energy minima.
\end{abstract}

\pacs{68.03.Cd, 47.85.-g, 89.90.+n, 83.80.Hj}

\maketitle

\section{Introduction}
Present day chemical synthesis methods allow for the creation of colloidal particles of various sizes, types, and shapes \cite{Kraft2009}. The behavior of these tailored particles in suspensions mimics that of atoms at favorable length and time scales, thus providing model systems for testing basic theoretical concepts, which even permit direct visualizations of thermal fluctuations \cite{Aarts2004}. In addition colloidal particles offer very rich perspectives for potential application, e.g., in the context of optically active materials \cite{Emory1998}. Colloidal particles of intermediate wettability usually adsorb at fluid-fluid interfaces. This configuration is extremely stable against thermal fluctuations due to the very large free energy of adsorption for micron-size particles. This can be used, for example, to stabilize emulsions \cite{Aveyard2003}. Such effectively two-dimensional colloidal systems have also been used to successfully test \cite{Zahn2000} the Kosterlitz-Thouless~\cite{Kosterlitz1973} scenario for 2D-melting, which is very difficult to capture in atomic systems. The presence of an interface affects effective interactions between colloidal particles considerably as compared to those for colloids in bulk solution. Examples for these interface induced effective interactions include capillary forces or effective dipole-dipole repulsive forces at water-air interfaces~\cite{HURD1985}. Whereas the gravity induced former ones do not play a role for colloids with a size below a micron, the latter ones explain the emergence of 2D-crystal structures~\cite{Pieranski1980} for laterally confined systems. However, there are experimental evidences ~\cite{Ruiz-Garcia1998,Ghezzi2001,Nikolaides2002} supporting the occurrence of attractive forces between equally charged particles at interfaces, with a range exceeding by far the range of van der Waals attraction. The observed attraction was originally attributed to flotation-like capillary forces decaying proportional to $d^{-1}$, where $d$ is the distance between the particles \cite{Nikolaides2002}. However, this explanation was found to be unsatisfactory~\cite{Oettel2005}, because for a mechanically isolated system, composed of the particles and the interface, one rather obtains an attractive behavior $\propto d^{-4}$ ~\cite{Dominguez2007a}. Consequently, the effective capillary attraction cannot overcome the direct dipole-dipole repulsion $\propto d^{-4}$, unless the screening length in water is comparable with the size of the particles. In this case a total effective interaction potential exhibits a shallow minimum with a depth of several $k_B T$~\cite{Dominguez2007a}. 
In Ref.~\cite{Oettel2005} it has been suggested that the finite curvature and the pinning of the interface might be important for generating a long-ranged attraction $\propto d^{-1}$. In fact, in the experiment reported in Ref.~\cite{Nikolaides2002} the particles were trapped at the surface of a droplet pinned to a solid plate, but this aspect was neglected in the theoretical analysis put forward in Ref.~\cite{Nikolaides2002}. 

\begin{figure}
\psfragscanon
\psfrag{substr}[c][c][1.5]{substrate}
\includegraphics[width=\textwidth]{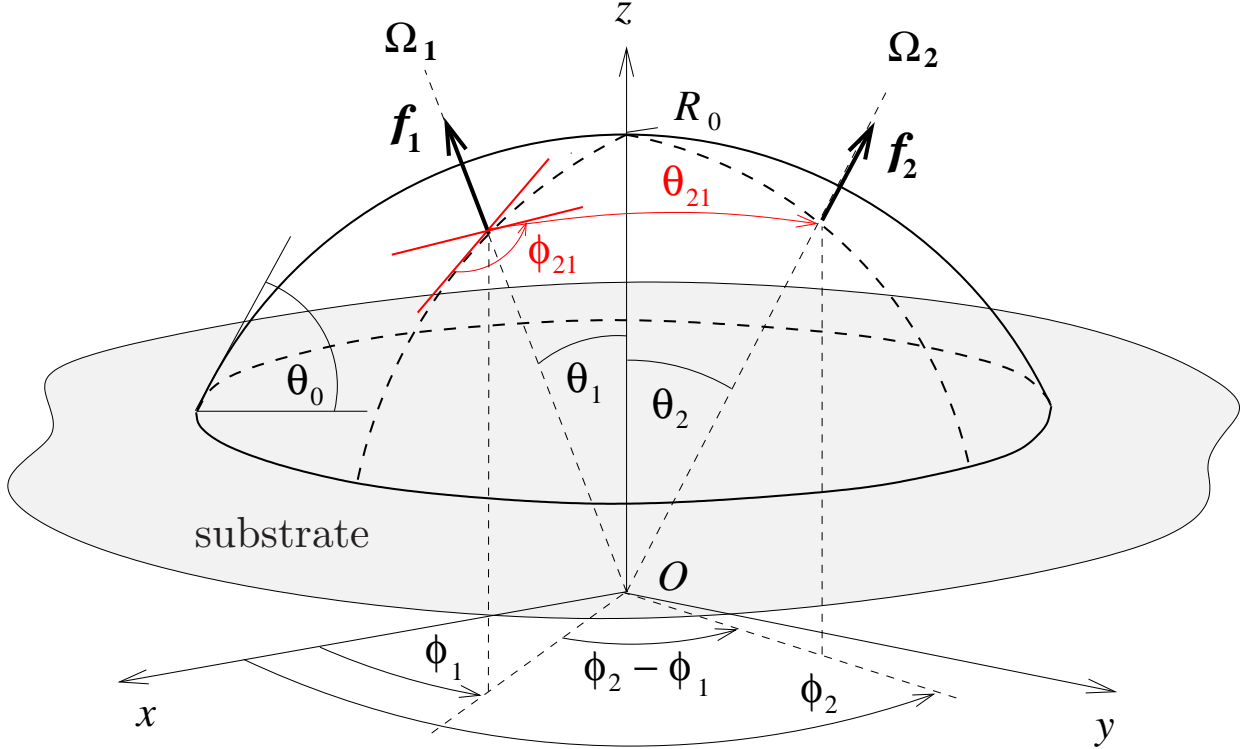}
\psfragscanoff
\caption[Spheres]{(Color online) Geometry of the system under consideration. The positions of the colloidal particles trapped at the surface of a sessile droplet with contact angle $\theta_0$ are parameterized by solid angles $\Omega_1$ and $\Omega_2$. These particles are subjected to radial external forces ${\bf f}_1=f_1{\bf \hat{e}}_r(\Omega_1)$ and ${\bf f}_2=f_2{\bf \hat{e}}_r(\Omega_2)$, respectively, where ${\bf \hat{e}}_r$ is the unit vector normal to the spherical cap-like surface of the droplet in the reference configuration. In Ref.\ \cite{Guzowski2010} it has been shown that under these conditions the particles can be well approximated by capillary monopoles, which in turn are determined exclusively by the external forces. This implies that the sizes of the particles and the contact angles at their surfaces are irrelevant. The dashed lines on the surface of the cap of the sphere with origin $O$ and radius $R_0$ are the points with spherical coordinates $(R_0,\theta_i,\phi_i)$, $i=1,2$, with $\phi_i$ fixed. $\theta_{21}$ is the angle formed by the vectors ${\bf f}_1$ and ${\bf f}_2$. The angle $\phi_{21}$ is defined in that tangential plane to the spherical cap for which ${\bf f}_1$ is a normal vector. At position $1$ two curves meet: one is the intersection of the plane $\phi_1=const$ through the origin $O$ with the spherical cap (dashed line), the other is the intersection of the spherical cap with the plane through $O$ spanned by ${\bf f}_1$ and ${\bf f}_2$ (long red arrow); $\phi_{21}$ is the angle formed by the vectors tangential to these two curves. 
\label{fig:system_sketch}
}
\end{figure}

Recently, it has been shown~\cite{Guzowski2010} that finite curvature and pinning indeed have profound effects on the trapping potential of a colloid as compared to the case of a flat and unbounded interface. In addition, general expressions have been obtained for the free energy of the system in the case of an arbitrary external pressure field acting on the droplet surface. In the present analysis we use those results in order to calculate the effective pair potential for particles subjected to external radial forces (see Fig.\ \ref{fig:system_sketch}).

We consider $N$ spherical particles floating at the surface of a sessile, non-volatile droplet, the surface of which meets the planar substrate at a contact angle $\theta_0$. With no external forces acting on the particles the shape of the droplet remains that of an undeformed spherical cap of radius $R_0$ and the immersions of the particles into the liquid forming the droplet are determined by the contact angles at each of the particle surfaces~\cite{Kralchevsky_book}. (Due to the partial immersion of the colloidal particles and the volume constraint of the liquid, the radius $R_0$ of the spherical cap with colloids is larger than that of the corresponding droplet without colloids.) We call this the reference configuration. We place the origin of the reference system at the geometrical center of a sphere formed by completing the unperturbed droplet shape to a full sphere, and choose the outward normal of the exposed substrate surface as the positive $z-$direction. We parametrize the positions of the particles by the radial distances $r_i$ and by the solid angles $\Omega_i=(\theta_i,\phi_i)$, where $\Omega=(\theta,\phi)$ is the solid angle used to parameterize the droplet surface (see Fig.~\ref{fig:system_sketch}).
In the reference configuration the particles float freely at the surface of the droplet, because in this case there is no preferred angular configuration of the particles. In the presence of radial external forces $f_i{\bf \hat{e}}_r$ (e.g., generated by optical tweezers) acting on each of the particles the droplet deforms such that those forces are balanced by the radial components of the corresponding capillary forces. However, the tangential capillary forces acting on the particles are in general unbalanced which leads to effective interactions between the particles. In terms of the free energy this means that in this case the total free energy depends on the angular configuration of the particles. 


\section{Formulation of the model in terms of capillary monopoles, Green's functions, and the free energy}

Previously it has been shown~\cite{Guzowski2010} that the influence of a single colloidal
particle on the surface  of a sessile droplet can be described in terms of a capillary 
monopole determined by the external radial force acting on the particle. The results of this approximate, linearized theory agree very well with the corresponding results of the numerical minimization of the full, non-Gaussian surface free energy. Therefore in the following we describe the particles in terms of capillary monopoles, so that the particle sizes and contact angles at their surfaces remain intrinsic parameters which do not enter the analysis explicitly. Accordingly, we study the droplet under the action of radial external pointlike forces of magnitudes $f_i$ giving rise to an external pressure $\Pi(\Omega) = \sum_{i=1}^{N}f_i\delta(\Omega,\Omega_i)/R_0^2$ with $\delta(\Omega,\Omega^{\prime}) = \delta(\theta-\theta^{\prime})\delta(\phi-\phi^{\prime})/\sin\theta$ as Dirac's angular delta function.

We consider the case that the deformations $u(\Omega)=R(\Omega)-R_0$ of the droplet surface $R(\Omega)$ due to the external pressure $\Pi(\Omega)$ are small. For this to hold, the condition~\cite{Guzowski2010} $|f_i|\ll\gamma R_0$, $i=1,\ldots,N$, where $\gamma$ is the tension of the droplet surface, might not be sufficient, because the forces on individual particles, even if they are small, might add up to a large total external force (and thus might lead to a large deformation); therefore we impose the more stringent condition $\sum_{i=1}^{N}|f_i|\ll\gamma R_0$. If this condition holds, $\epsilon:=\sum_{i=1}^{N}|f_i|/(\gamma R_0)$ can be regarded as a small parameter. (In the one-particle case this reduces to $\epsilon=f/(\gamma R_0)$ as introduced in Ref.~\cite{Guzowski2010}). On the other hand one has to keep in mind that the forces $f_i$ cannot be stronger than the maximal capillary forces, which lead to the extraction of the individual particles from the surface, and which are of the order of $\gamma a_i$, where $a_i$ is the characteristic size of the particle $i$. 

The resulting shape of the droplet can be determined by minimizing a free energy functional under the constraint of a fixed liquid volume, which leads to the Young-Laplace equation relating the curvature of the droplet surface to the pressure difference across the interface~\cite{Guzowski2010}. This equation depends also on the mechanism fixing the lateral position of the droplet at the substrate. In the following we shall use the subscript $\sigma$ in order to distinguish between the cases of a free contact line with a fixed center of mass ($\sigma=A$) and of a pinned contact line with a free center of mass ($\sigma=B$; in this case the center of mass can be free because the balance of forces is already guaranteed by the pinning of the contact line). Linearized in $\epsilon$ the Young-Laplace equation, expressed in terms of the dimensionless deformation $v:=\lim_{\epsilon\rightarrow0}u/(R_0\epsilon)$, takes the form (see Eq.\ (29) in Ref.~\cite{Guzowski2010})
\begin{equation}
	-(\nabla_a^2+2)v(\Omega)=\pi(\Omega) + \pi_{CM}(\Omega) + \mu,
	\label{helmholtz}
\end{equation}
\noindent where $\nabla_a:={\bf e}_{\theta}\partial_{\theta} +\dfrac{{\bf e}_{\phi}}{\sin\theta}\partial_{\phi}$ is the dimensionless {\em a}ngular gradient on the unit sphere and $\mu$ is the spatially uniform dimensionless shift of the internal droplet pressure relative to the Laplace pressure $2\gamma/R_0$ of a spherical droplet. This shift occurs due to the external excess surface pressure $\pi(\Omega):=\lim_{\epsilon\rightarrow0}\Pi(\Omega)R_0/(\gamma\epsilon)$ exerted by the external forces $f_i$. The free energy approach reveals that the internal pressure is equal to $\lambda$, where $-\lambda$ is the Lagrange multiplier introduced into the free energy functional in order to fix the liquid volume. Therefore $\mu$ can be identified as 
$\mu=\lim_{\epsilon\rightarrow0}(\lambda(\epsilon)-2\gamma/R_0)R_0/(\gamma\epsilon)$. 
The effective pressure $\pi_{CM}(\Omega)$ corresponds to a body force fixing the center of mass of the droplet, which has to be introduced in the case of a free contact line $(\sigma=A)$ in order to achieve mechanical equilibrium; accordingly, it depends on the total external force $\sum_{i=1}^{N}f_i{\bf e}_{i}(\Omega_i)$. In the case of a free center of mass $(\sigma=B)$ one has $\pi_{CM}=0$. Equation (\ref{helmholtz}) is supplemented by the boundary condition $\sin\theta_0\partial_{\theta}v|_{\theta_0}-\cos\theta_0v|_{\theta_0}  = 0$ for the case of a free contact line and $v\vert_{\theta_0}=0$ for the case of a pinned contact line (Eqs.\ (32) and (33) in Ref.~\cite{Guzowski2010}) and by the incompressibility condition 
\begin{equation}
	 \int_{\Omega_0}\!  d\Omega\, v(\Omega) = 0,
\label{incopress}
\end{equation}
where the angular domain $\Omega_0$ corresponds to the shape of the reference droplet, which is a spherical cap with the contact angle $\theta_0$ at the substrate.

The formal solution $v\equiv v_{\sigma}$ of Eq.\ (\ref{helmholtz}) can be written as
\begin{equation}
	v_{\sigma}(\Omega)=\int_{\Omega_0}\! d\Omega'\, \Pi(\Omega')G_{\sigma}(\Omega,\Omega',\theta_0),
	\label{AB-solution}
\end{equation}
where Green's functions $G_{\sigma}$ satisfy
\begin{equation}
	-(\nabla_a^2+2)G_{\sigma}(\Omega,\Omega',\theta_0)
	=\delta(\Omega,\Omega')+\Delta_{\sigma}(\Omega,\Omega',\theta_0).
\label{govG2}
\end{equation}
The functions $\Delta_{\sigma}(\Omega,\Omega',\theta_0)$ correspond to $\mu$ and $\pi_{CM}$ in Eq.\ (\ref{helmholtz}), so that $\mu+\pi_{CM}(\Omega)=\int_{\Omega_0}\!d\Omega'\, \pi(\Omega')\Delta_{\sigma}(\Omega,\Omega',\theta_0)$. They can be determined from the force balance and the incompressibility condition~\cite{Guzowski2010}. Here, we only point out that, in general, these functions are not symmetric with respect to interchanging $\Omega$ and $\Omega'$ and so neither are Green's functions $G_{\sigma}(\Omega,\Omega',\theta_0)$. The dependence on $\sigma$ enters via the boundary conditions for $G_{\sigma}$:
\begin{align}
 \sin\theta_0\partial_{\theta}G_A(\Omega,\Omega',\theta_0)\vert_{\Omega\in\partial\Omega_0}-\cos\theta_0G_A(\Omega,\Omega',\theta_0)\vert_{\Omega\in\partial\Omega_0} &= 0, \label{boundary_N}\\
 G_B(\Omega,\Omega',\theta_0)\vert_{\Omega\in\partial \Omega_0} &= 0. \label{boundary_D}
\end{align}
The incompressibility condition can be expressed as
\begin{equation}
	\int_{\Omega_0} d\Omega\, G_{\sigma}(\Omega,\Omega',\theta_0) = 0,
	\label{volume_ND}
\end{equation}
see Eqs.\ (50)-(52) in Ref.~\cite{Guzowski2010}. We point out that, locally, i.e., for $\Omega\rightarrow\Omega'$, Eq.\ (\ref{govG2}) can be approximated as $-\nabla_a^2G_{\sigma}(\Omega,\Omega',\theta_0)=\delta(\Omega,\Omega')$, from which it follows that $G_{\sigma}(\Omega,\Omega',\theta_0)\xrightarrow[\Omega\rightarrow\Omega']{}-\ln(\bar{\theta})/(2\pi)$ where $\bar{\theta}$ is the angle between the unit vectors pointing into the directions $\Omega$ and $\Omega'$ (see Eq.\ (57) in Ref.~\cite{Guzowski2010}). By adopting the analogy with $2D$ electrostatics~\cite{Morse1993,Dominguez2008a}, this logarithmic divergence can be traced back to the diverging self-energy $F_{self}$ of a monopole associated with a pointlike force. However, in an actual system the size $a$ of the particle serves as a natural cutoff for the divergence of the self-energy which for $a\rightarrow0$ scales as $F_{self}\sim f^2\ln(R_0/a)\lesssim \gamma^2 a^2\ln(R_0/a)$, using $f\lesssim \gamma a$.

The excess free energy associated with the droplet deformation due to the external pressure
$\pi(\Omega)$ can be written in the form (compare with Eq.\ (53) in Ref.~\cite{Guzowski2010})
\begin{equation}
	F_{\sigma} = - \dfrac{1}{2}\epsilon^2 \gamma R_0^2\int_{\Omega_0} \!d\Omega\, \int_{\Omega_0} \!d\Omega'\, \pi(\Omega)G_{\sigma}(\Omega,\Omega',\theta_0)\pi(\Omega').
\label{free_energy}
\end{equation}
Inserting $\pi(\Omega) = \sum_{i=1}^{N}f_i\delta(\Omega,\Omega_i)/(\sum_{i=1}^{N}|f_i|)$ one obtains
\begin{equation}
	F_{\sigma} = F_{\sigma}^{(N)} = \sum_{i=1}^N F_{i, self} + \sum_{i=1}^N \Delta F^{(1)}_{\sigma,i}(\theta_i,\theta_0) +\sum_{i<j}V_{\sigma}(\Omega_i,\Omega_j),
\label{free_energy2}
\end{equation}
where $F_{i, self}= f_i^2/(4\pi\gamma)\ln(R_0/a_i)+O(1)$ is the self-energy of particle $i$, which depends neither on the position of the particle on the droplet nor, in leading order in $a_i/R_0$, on the contact angle $\theta_0$ and the boundary conditions $\sigma$. The one-particle free energy landscapes $\Delta F^{(1)}_{\sigma,i}$ are given by 
\begin{equation}
	\Delta F^{(1)}_{\sigma,i}(\theta_i,\theta_0) =-\frac{f_i^2}{2\gamma}[G_{\sigma,reg}(\Omega_i,\Omega_i,\theta_0)-G_{\sigma,reg}(0,0,\theta_0)]
	= -\frac{f_i^2}{2\gamma}[g_{\sigma}(\theta_i,\theta_0)-g_{\sigma}(0,\theta_0)], \label{excess_free_enND}
\end{equation}
where the function $g_{\sigma}$ is independent of $i$ and corresponds to the regular part $G_{\sigma,reg}$ (i.e., not containing the logarithmic divergence) of Green's function which can be found analytically in the case $\theta_0=\pi/2$ (see Sec.\ IV.C in Ref.~\cite{Guzowski2010} and Fig.\ \ref{fig:gsigma}) and numerically otherwise~\cite{Guzowski2010}. The last term in Eq.\ (\ref{free_energy2}) consists of effective pair potentials $V_{\sigma}$ given by
\begin{equation}
	V_{\sigma}(\Omega_i,\Omega_j,\theta_0)=-\dfrac{f_if_j}{2\gamma}\big[G_{\sigma}(\Omega_i,\Omega_j,\theta_0)+G_{\sigma}(\Omega_j,\Omega_i,\theta_0)\big].
\label{pair-pot}
\end{equation}
We note that $V_{\sigma}$ is explicitly symmetric with respect to $\Omega_i$ and $\Omega_j$ even if $G_{\sigma}$ is not.

For the special case $\theta_0=\pi/2$ one can construct $G_{\sigma}$ by using the method of images from electrostatics  \cite{Guzowski2010} (see Sec.\ IV in Ref.~\cite{Guzowski2010}). In the reference configuration the substrate is replaced by a virtual mirror image of the actual droplet, such that the union of the actual drop and of the virtual drop forms a full spherical droplet. The deformations of a fully spherical drop have been studied by Morse and Witten~\cite{Morse1993}, who derived the following equation for a ``free'' Green's function $G$: 
\begin{equation}
	-(\nabla_a^2+2)G(\Omega,\Omega')=\sum_{l\geq 2,m}Y^*_{lm}(\Omega)Y_{lm}(\Omega')=\hat{\delta}(\Omega,\Omega'),
	\label{greens}
\end{equation}
\noindent where the right hand side is an expansion in terms of spherical harmonics of a modified Dirac delta function $\hat{\delta}(\Omega,\Omega')$ with the $l=0$ and $l=1$ components projected out. Summing up the coefficients in the expansion in terms of spherical harmonics one obtains the solution in closed form:
\begin{equation}
G(\Omega,\Omega')\equiv G(\bar{\theta})=-\frac{1}{4\pi}\left[ \frac{1}{2} + \frac{4}{3}\cos\bar{\theta} + \cos\bar{\theta}\ln\left(\frac{1-\cos\bar{\theta}}{2}\right) \right].
\label{greens_function}
\end{equation}
In the case of a sessile drop the solution is constructed in the form of a superposition of the point-force solution given by Eq.~(\ref{greens_function}) (which would be valid, if there was no substrate) and a solution associated with a virtual pointlike force acting symmetrically at the surface of the virtual lower hemisphere. For constructing Green's functions one has to take into account the boundary conditions as well as the mechanisms imposing the corresponding force balance. In the case of a free contact line the droplet is not attached to the substrate and mechanical equilibrium is achieved by fixing the lateral position of the center of mass of the liquid. This latter aspect is already incorporated into Green's function as given by Eq.\ (\ref{greens}). However, if the balance of forces is imposed by a pinned contact line and the center of mass is not fixed, the corresponding theoretical treatment requires the introduction  of additional images, so that the total force vanishes. Concerning these details we refer the reader to Refs.~\cite{Guzowski2010} and~\cite{Morse1993}.

\section{Results}
\subsection{Contact angle $\theta_0=\pi/2$}
\begin{figure}[ht]
\includegraphics[width=0.6\textwidth]{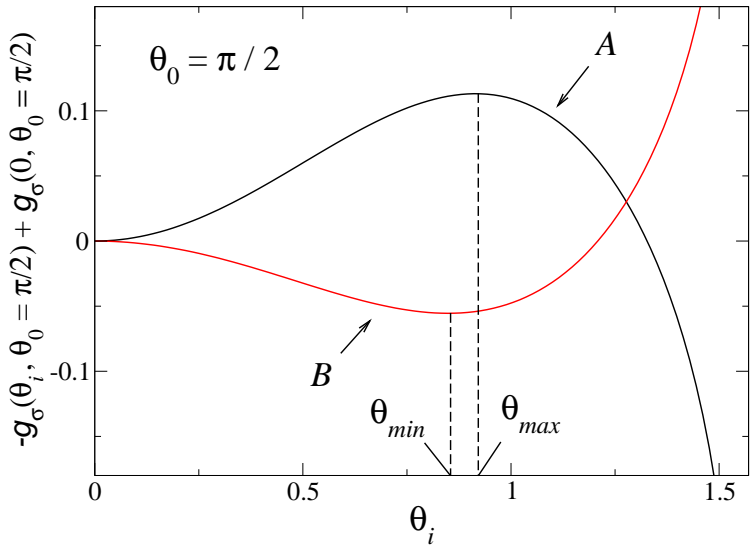}
\caption[]{The functions $-g_{\sigma}(\theta_i,\theta_0=\pi/2)+g_{\sigma}(0,\theta_0=\pi/2)$, $\sigma=A,B$, which determine the behavior of the one-particle free energy landscape $\Delta F^{(1)}_{\sigma,i}(\theta_i,\theta_0)$ of particle $i$, as functions of its polar angle $\theta_i$, for $\theta_0=\pi/2$ (see Eq.\ (\ref{excess_free_enND})).
 \label{fig:gsigma}
}
\end{figure}
\begin{figure}
\begin{overpic}[width=\textwidth]{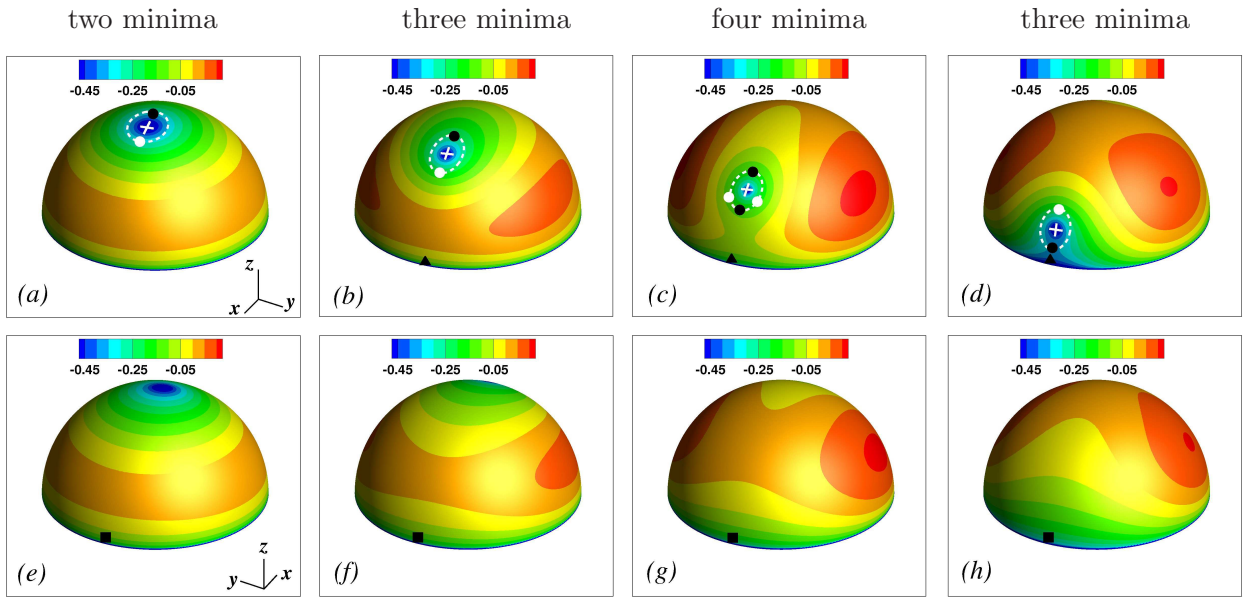}
 \put(5,46){two minima}
 \put(32,46){three minima}
 \put(57,46){four minima}
 \put(82,46){three minima}
\end{overpic}
\caption[]{Color coded effective pair potential for the case $\theta_0=\pi/2$
  and for a free contact line at the substrate ($\sigma=A$). The colors
  represent the excess surface free energy $\gamma\Delta F^{(2)}_{A}/(f_1f_2)$
  (Eqs.\ (\ref{free_energy2}) and (\ref{excess_free_energy})) of the system as
  a function of the position $\Omega_2$ of the probe particle for a fixed
  position $\Omega_1$ of the reference particle (white cross). Each pair of panels in a column corresponds to the same configuration.
The white
  dashed lines schematically indicate those positions $\Omega_2$ of the probe
  particle which are at a fixed minimal polar angular separation
  $\theta_{21}=\delta=\pi/36$ from the reference particle, i.e., at mutual
  contact. Along these dashed lines $\Delta F_A^{(2)}$ is locally minimal
  (maximal) at the black (white) dots. The black symbols
  (\ding{108},\ding{110},\ding{115}) correspond to minima under the
  constraints $\theta_{21}>\delta$ and $\theta_2<\pi/2-\delta$.
\textit{(a), (e)}[back side]:  $\Omega_1=(\theta_1=\pi/18,\phi_1=0)$;
\textit{(b), (f)}[back side]:  $\Omega_1=(\theta_1=3\pi/18,\phi_1=0)$;
\textit{(c), (g)}[back side]:  $\Omega_1=(\theta_1=5\pi/18,\phi_1=0)$;
\textit{(d), (h)}[back side]:  $\Omega_1=(\theta_1=7\pi/18,\phi_1=0)$.
Note that in $(d)$ two minima are shown (\ding{108},\ding{115}) separated by a potential ridge.
In \textit{(c)} the value $\theta_1=5\pi/18=\pi/3.6$ is taken slightly smaller than $\theta_{max}\approx \pi/3.46$, which is the position of the free energy maximum for the reference particle alone (see Fig. \ref{fig:gsigma}). In such a case the upper black dot marks the deeper one of the two local minima on the dashed curve. The two local maxima (white dots) are equally high. The orientations of the coordinate axes in \textit{(b)-(d)}[\textit{(f)-(h)}] are the same as in \textit{(a)}[\textit{(e)}]. The quoted numbers of minima are the total ones occurring for the given configurations shown in the columns of panels. 
 \label{fig:Model_A}
}
\end{figure}
\begin{figure}
\begin{overpic}[width=0.75\textwidth]{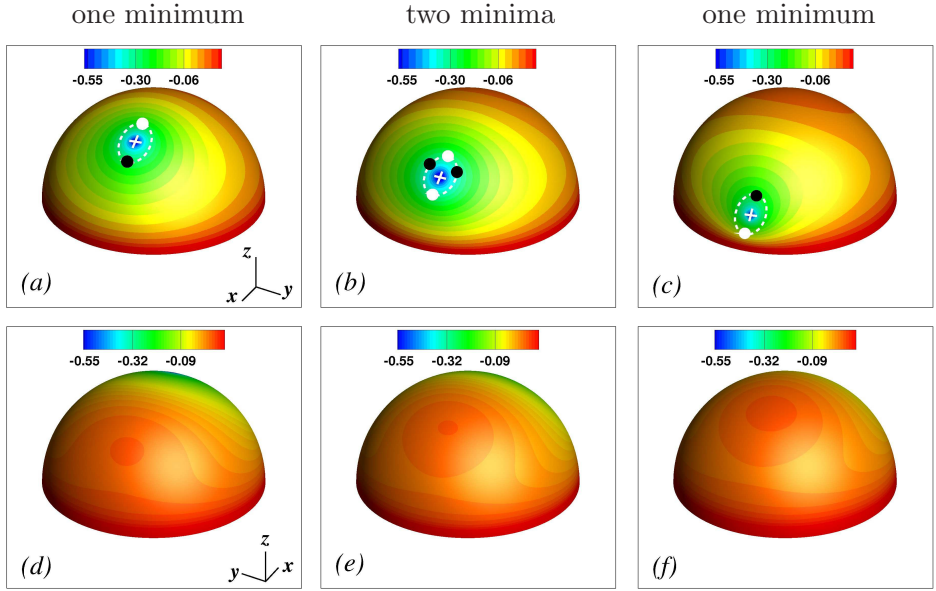}
 \put(7,61){one minimum}
 \put(43,61){two minima}
 \put(75,61){one minimum}
\end{overpic}
\caption[]{Same as Fig.\ \ref{fig:Model_A} for a pinned contact line at the substrate ($\sigma=B$). 
\textit{(a), (d)}[back side]:  $\Omega_1=(\theta_1=3\pi/18,\phi_1=0)$;
\textit{(b), (e)}[back side]:  $\Omega_1=(\theta_1=5\pi/18,\phi_1=0)$;
\textit{(c), (f)}[back side]:  $\Omega_1=(\theta_1=7\pi/18,\phi_1=0)$.
In \textit{(b)} the value $\theta_1=5\pi/18=\pi/3.6$ is taken almost equal to $\theta_{min}\approx \pi/3.67$ which is the position of the free energy minimum for the reference particle alone (see Fig.\ \ref{fig:gsigma}). 
The positions $\Omega_1$ (white crosses) in \textit{(a), (b)}, and \textit{(c)} equal those in Figs.\ \ref{fig:Model_A}\textit{(b), (c)}, and \textit{(d)}, respectively.
In \textit{(b)} the two minima on the dashed white curve are degenerate. The lower white dot marks the higher local maximum on the dashed white curve.
\label{fig:Model_B}
}
\end{figure}

First, we present our analytical results for the case $\theta_0=\pi/2$. As can be inferred from Eq.\ (\ref{excess_free_enND}) and Fig.\ \ref{fig:gsigma} the one-particle free energy landscape $\Delta F^{(1)}_{\sigma, i}(\theta_i,\theta_0=\pi/2)$ of particle $i$ is a non-monotonic function of its polar angle $\theta_i$. As a consequence, besides the known phenomena of attraction of a particle to a free contact line (model $A$) and repulsion from a pinned one (model $B$), one finds a local free energy minimum for the particle being at the drop apex $\theta_i=0$ and a local maximum at $\theta_i=\theta_{max}\approx52^{\circ}$ for model $A$, and a minimum at $\theta_i=\theta_{min}\approx49^{\circ}$ for model $B$. In the case of many particles the one-particle free energy landscapes compete with the effective pair potentials $V_{\sigma}$. The quantity relevant for obtaining the actual configurations of the particles is the excess free energy defined as
\begin{equation}
	\Delta F_{\sigma}^{(N)}:=  F^{(N)}_{\sigma} - \sum_{i=1}^N F_{i, self} 
	= \sum_{i=1}^N \Delta F^{(1)}_{\sigma, i}+\sum_{i<j}V_{\sigma,ij},
\label{excess_free_energy}
\end{equation}
where we have introduced the notation $V_{\sigma,ij}\equiv V_{\sigma}(\Omega_i,\Omega_j,\theta_0)$. In Figs.\ \ref{fig:Model_A} and \ref{fig:Model_B} we present the results for $\Delta F_{\sigma}^{(N)}$ in the case $N=2$. We consider particles placed at $\Omega_1=(\theta_1,\phi_1)$ and $\Omega_2=(\theta_2,\phi_2)$ and subjected to external radial forces $f_1$ and $f_2$ such that $|f_1|\ll\gamma R_0$, $|f_2|\ll\gamma R_0$, and $|f_1+f_2|\ll\gamma R_0$. The excess free energy $\Delta F_{\sigma}^{(2)}$ is calculated for a fixed angular position of the first particle, referred to as the reference particle, as a function of the angular position of the second one acting as a probe particle. The angular cutoff $\delta$ determines the closest approach of the particles to each other, i.e., $\theta_{21}>\delta$, and to the contact line, i.e., $\theta_1,\theta_2\in[0,\theta_0-\delta]$, where $\theta_{21}$ is the angle between the vectors pointing in the directions $\Omega_1$ and $\Omega_2$ (see Fig.\ \ref{fig:system_sketch}). One has $\delta\gtrsim 2a/R_0$, but one has to remember that for the configurations corresponding to $\theta_{21}\simeq\delta$ one should expect that the actual behavior deviates from that obtained within the monopole approximation. In the following calculations we take $\delta=\pi/36$, which corresponds to $R_0\simeq23a$.

\begin{figure}
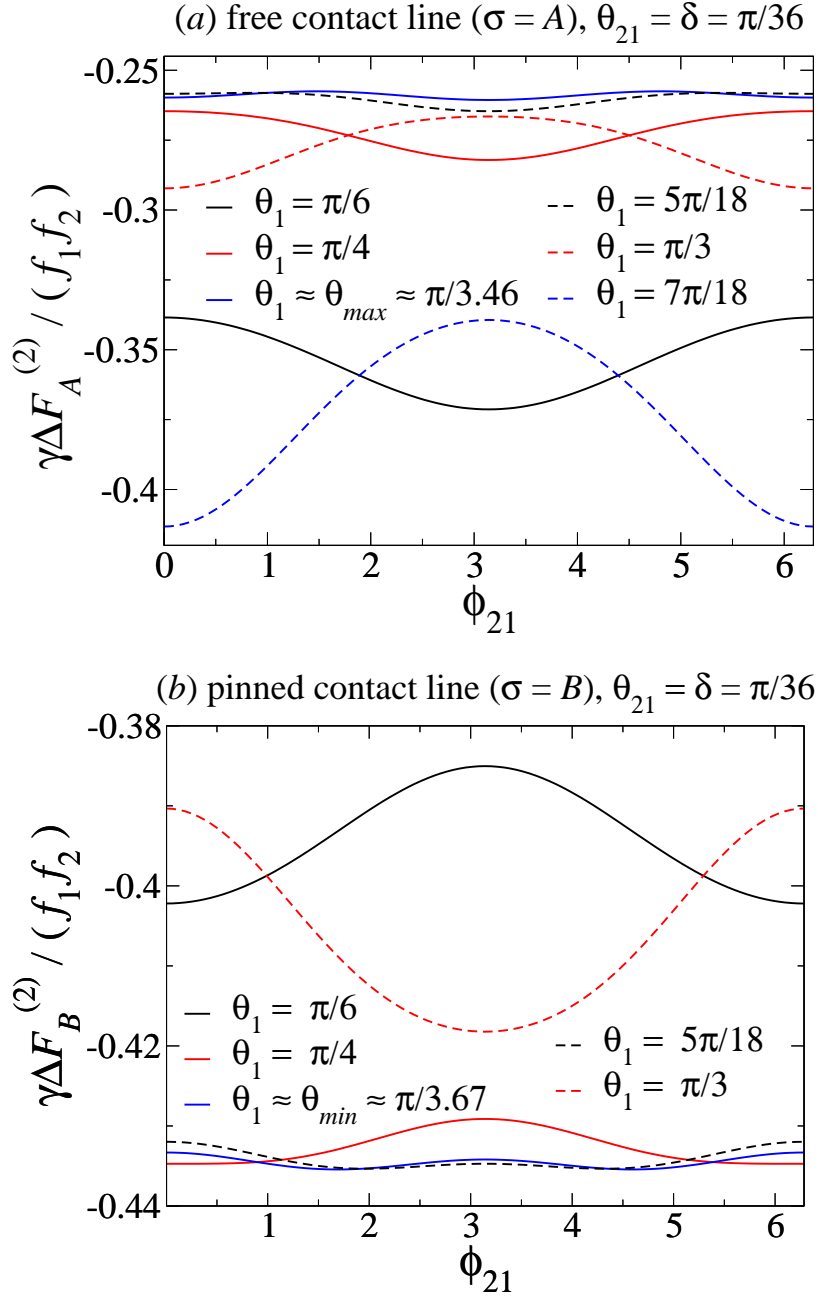

\begin{overpic}[width=0.65\textwidth]{fig05}
\end{overpic}\\
\vspace*{0.5cm}
\begin{overpic}[width=0.65\textwidth]{fig06}
\end{overpic}
\caption[]{Normalized free energy $\Delta F_{\sigma}^{(2)}$
  (Eqs.\ (\ref{free_energy})-(\ref{pair-pot}) and (\ref{excess_free_energy}))
  of a pair of colloidal particles floating on a sessile droplet with contact
  angle $\theta_0=\pi/2$ and exposed to external radial forces $f_i$ as a
  function of the angle $\phi_{21}$ (see Fig.~\ref{fig:system_sketch}) for
  several fixed values of $\theta_1$ characterizing
the polar angular position of the reference particle. Concerning $\theta_{max}$ and $\theta_{min}$ see Fig.\ \ref{fig:gsigma}. The polar angular
separation $\theta_{21}$ (see Fig.\ \ref{fig:system_sketch}) between two
particles is kept constant upon varying $\phi_{21}$ and equals
$\theta_{21}=\delta=\pi/36$, i.e., the free energy landscape is probed around
the reference particle fixed in space. $(a)$ free contact line; $(b)$ pinned
contact line.
 \label{fig:free_energy_vs_phi21}
}
\end{figure}

\begin{figure}
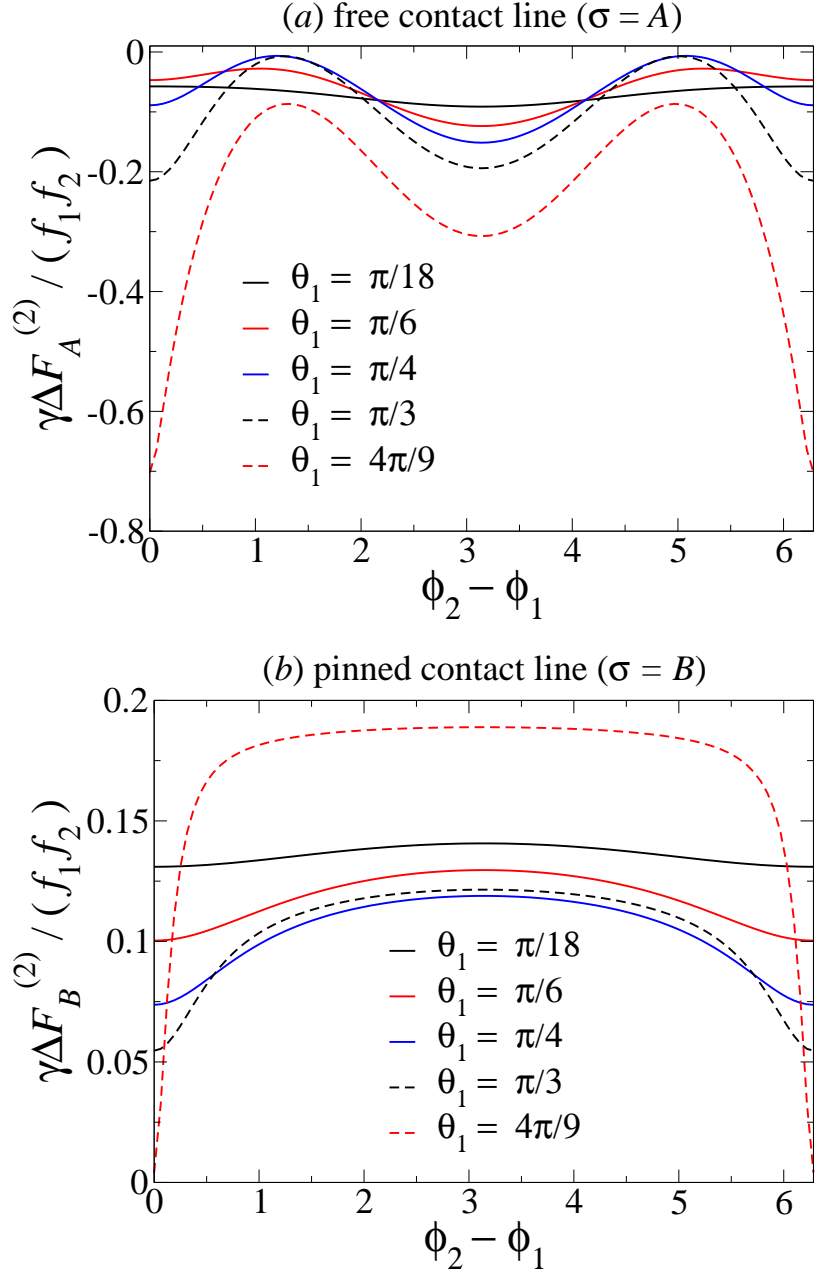

\begin{overpic}[width=0.65\textwidth]{fig07}
\end{overpic}\\
\vspace*{0.5cm}
\begin{overpic}[width=0.65\textwidth]{fig08}
\end{overpic}
\caption[]{Normalized free energy (Eqs.\ (\ref{free_energy})-(\ref{pair-pot}) and (\ref{excess_free_energy})) of a pair of colloidal particles floating on a sessile droplet with contact angle $\theta_0=\pi/2$ and exposed to external radial forces $f_i$
as a function of the relative azimuthal angle $\phi_{2}-\phi_{1}$ (see Fig.\ \ref{fig:system_sketch})) for several fixed polar angles $\theta_1$ of the reference particle. Here, the free  energy landscape is probed 
close to the substrate, i.e., $\theta_2 = \theta_0-\delta = \pi/2 - \pi/36$. $(a)$ free contact line; $(b)$ pinned contact line.
 \label{fig:free_energy_vs_phi2-phi1}
}
\end{figure}

The interaction potential $V_{\sigma}$ is a function of the angular coordinates of both particles separately, i.e., it is not only a function of their separation. This is important for all angular configurations due to the long range of the capillary deformation around a monopole. This deformation does not depend on the radius $R_0$ of the droplet but only on the strength of the external force. In those cases in which the particles would correspond to higher capillary multipoles the pair interactions would vanish for $R_0\rightarrow\infty$; in this sense the case of monopoles is exceptional. We also note that, as given by Eq.\ (\ref{pair-pot}), $V_{\sigma}$ is explicitly symmetric with respect to $\Omega_1$ and $\Omega_2$, such that the aforementioned asymmetry of Green's functions $G_{\sigma}$ is not proliferated to the free energy.

Due to the rotational symmetry of the reference droplet forming a spherical cap the free energy depends only on the difference $\phi_2-\phi_1$ so that $\Delta F^{(2)}_{\sigma}=\Delta F^{(2)}_{\sigma}(\theta_1,\theta_2,\phi_1-\phi_2)$. We also introduce an auxiliary azimuthal angle $\phi_{21}$ which describes the angular position of the probe particle relative to the reference one (see Fig.\ \ref{fig:system_sketch}) and thus the orientation of the pair of particles on the droplet.

In the case of a {\em free contact line} (Fig.\ \ref{fig:Model_A}) at the substrate and $\theta_1=0$ the minima of the free energy are degenerate both for the configurations with the probe particle at $\theta_2=\delta$ (i.e., in contact with the reference particle) and for those with $\theta_2=\pi/2-\delta$ (i.e., the probe particle being in contact with the substrate). In both cases the dependence on $\phi_2$ is degenerate. However, if $\theta_1\neq0$, the rotational symmetry is broken and this degeneracy is lifted.  The precise values of the angular positions of the free energy minima depend on the value of $\theta_1$ but in general they correspond either to the closest approach of the particles to each other or to the contact line. If $\theta_1\ll 1$, there are only two free energy minima. They occur at $\theta_{21}=\delta$, $\phi_{21}=\pi$ [\ding{108}] (see $\theta_1=\pi/6$ in Fig.\ \ref{fig:free_energy_vs_phi21}$(a)$) and at $\theta_2=\pi/2-\delta$, $\phi_2-\phi_1=\pi$ [\ding{110}] (see $\theta_1=\pi/18$ in Fig.\ \ref{fig:free_energy_vs_phi2-phi1}$(a)$). Increasing $\theta_1$ leads first to the emergence of a third minimum at $\theta_2=\pi/2-\delta$, $\phi_2-\phi_1=0$ [\ding{115}] (see $\theta_1=\pi/6$ in Fig.\ \ref{fig:free_energy_vs_phi2-phi1}$(a)$) followed by the emergence of a fourth one at $\theta_{21}=\delta$, $\phi_{21}=0$ [\ding{108}] (see $\theta_1=\theta_{max}\approx52^{\circ}$ in Fig.\ \ref{fig:free_energy_vs_phi21}$(a)$; for $\theta_{max}$ compare Fig.\ \ref{fig:gsigma}). Finally, upon further increasing $\theta_1$ the minimum at $\theta_{21}=\delta$, $\phi_{21}=\pi$ turns into a local maximum, leaving three local minima at $(\theta_{21}=\delta,\phi_{21}=0)$ [\ding{108}], $(\theta_{2}=\pi/2-\delta,\phi_{2}-\phi_1=0)$ [\ding{115}], and at $(\theta_2=\pi/2-\delta,\phi_2-\phi_1=\pi)$ [\ding{110}] (see $\theta_1=\pi/3$ in Figs.\ \ref{fig:free_energy_vs_phi21}$(a)$ and \ref{fig:free_energy_vs_phi2-phi1}$(a)$). The answer to the question which configuration corresponds to the global free energy minimum with respect to the positions of {\em both} particles depends on $\delta$. For $\delta\ll1$ it is the configuration in which both particles touch the substrate and touch each other (see $\theta_1=4\pi/9$, $\phi_{2}-\phi_1=0$ in Fig.\ \ref{fig:free_energy_vs_phi2-phi1}$(a)$; for $\theta_0=\pi/2$ one has $\theta_i\leq\pi/2$), because the depth of the corresponding minimum increases as $\sim\ln(1/\delta)$. Thus the particles arrange themselves parallel to the contact line.

For a {\em pinned contact line} (Fig.\ \ref{fig:Model_B}) at the substrate we also observe a degenerate minimum of the free energy for $\theta_1=0$ and a broken symmetry for  $\theta_1\neq 0$. Upon further increasing $\theta_1$, first there is only a single free energy minimum, occurring at $\theta_{21}=\delta$, $\phi_{21}=0$ (see $\theta_1=\pi/6$ in Fig.\ \ref{fig:free_energy_vs_phi21}$(b)$). Upon increasing $\theta_1$ further this minimum splits continuously into two minima at $(\theta_{21}=\delta,\phi_{21}')$ with $\phi_{21}'\in[0,\pi]$ and at $(\theta_{21}=\delta,\phi_{21}'')$ with $\phi_{21}''\in[\pi,2\pi]$ (see Fig.\ \ref{fig:free_energy_vs_phi21}$(b)$ for $\theta_1\gtrsim\pi/4$). For $\theta_{1}=\theta_{min}$ the minima have reached the values $\phi_{21}'\approx\pi/2$ and $\phi_{21}''\approx3\pi/2$, which corresponds to the configuration in which the particles are positioned parallel to the contact line (see $\theta_1=\theta_{min}\approx 49^{\circ}$ in Fig.\ \ref{fig:free_energy_vs_phi21}$(b)$). Finally, for even larger $\theta_1$ the two minima merge into a single minimum at $\theta_{21}=\delta$, $\phi_{21}=\pi$ (see $\theta_1=\pi/3$ in Fig.\ \ref{fig:free_energy_vs_phi21}$(b)$). The capillary forces repel the particles from the contact line (see Fig.\ \ref{fig:gsigma}) and therefore all configurations with any of the particles close to the contact line are energetically unfavorable. However, it might happen that one of the particles gets trapped close to the contact line by other means, for example due to an evaporative flux of the liquid towards the contact line or by adhesion to the substrate. If the probe particle be the trapped one, such that $\theta_2=\pi/2-\delta$ and $\phi_2$ is free, and for any fixed polar position $\theta_1$ of the reference particle (see Fig.\ \ref{fig:free_energy_vs_phi2-phi1}$(b)$), the preferred position of the probe particle at the contact line always corresponds to $\phi_2-\phi_1=0$, i.e., the two particles are positioned at a great circle perpendicular to the contact line. This means that for any angular position of the white cross in Fig.\ \ref{fig:Model_B} the minimum of the free energy along the contact line occurs at the point closest to the white cross.
In the case that both particles are constrained to lie in the neighborhood of the contact line one observes a monotonic attraction (Fig.\ \ref{fig:free_energy_vs_phi2-phi1}$(b)$, $\theta_1=4\pi/9$), contrary to the case of a free contact line (Fig.\ \ref{fig:free_energy_vs_phi2-phi1}$(a)$, $\theta_1=4\pi/9$). The configuration corresponding to the global free energy minimum with respect to the positions of {\em both} particles without any constraints is such that the particles touch each other at $\theta_1=\theta_2\approx\theta_{min}\approx 49^{\circ}$ (see $\theta_1=\theta_{min}$, $\phi_{21}=\pi\pm\pi/2$ in Fig.\ \ref{fig:free_energy_vs_phi21}$(b)$). Thus the particles spontaneously arrange themselves parallel to the contact line at the common characteristic polar angle $\theta_{min}$. 

In summary, in both cases the particles attract each other and, as a doublet, arrange themselves such that they are both placed as close as possible to the minimum of the one-particle trapping potential $\Delta F^{(1)}_{\sigma}$ which, in the case of a free contact line, occurs at the apex and at the contact line whereas in the case of a pinned contact line it occurs at an intermediate angle $\theta_{min}$. Additionally, in the case of a free contact line there is another local free energy minimum corresponding to both particles being at the contact line, however not touching each other but being positioned on the opposite sides of the droplet ($\phi_2-\phi_1=\pi$).


\subsection{Arbitrary contact angle}

In the case of an arbitrary contact angle $\theta_0\neq\pi/2$ we minimize the free energy numerically by using a finite element method~\cite{Brakke}. 

First, the one-particle free energy landscapes $\Delta F^{(1)}_{\sigma,i}(\theta_i,\theta_0)$ are calculated for various values of the contact angle $\theta_0$ and the force $f_i$ acting on particle $i$. The numerical procedure is similar to the one described in Ref.\ \cite{Guzowski2010} and consists of a pre-evolution of a body of liquid towards the reference configuration ($f_i=0$) for a fixed polar angle $\theta_i$ of particle $i$, after which the force is turned on ($f_i\neq0$) and the droplet surface evolves towards the minimum energy while the particle is allowed to move only radially. We recover the scaling of the free energy $\sim f_i^2$, which yields the scaling functions $g_{\sigma}(\theta_i,\theta_0)$ (Eq.\ (\ref{excess_free_enND})). In the case of a free contact line we observe small deviations from this scaling, which we expect to vanish for very large droplets (which, however, are beyond the reach of the finite element method). There are two sources of this deviation. First, there is a finite-size effect due to fixing the center of mass of the droplet which leads to a contribution to the free energy associated with the work done by the virtual counterbalancing force $f_{CM,i}(\theta_i)=f_i\sin\theta_i$ in displacing the center of mass from its reference lateral position $x=x_{CM,ref}(\theta_i)$, which does not depend on $f_i$, to $x=0$ (for details see Ref.~\cite{Guzowski2010}). Thus, this correction is linear in $f_i$ and it can be eliminated by taking the average of the results for $f_i=|f_i|$ and $f_i=-|f_i|$. Second, there is a numerical error associated with the possibility that the evolving surfaces can get trapped in local free energy minima and thus do not reach the global free energy minimum. Performing the calculations for positive and negative values of $f_i$ and subsequent averaging minimizes this error, too. 

\begin{figure}
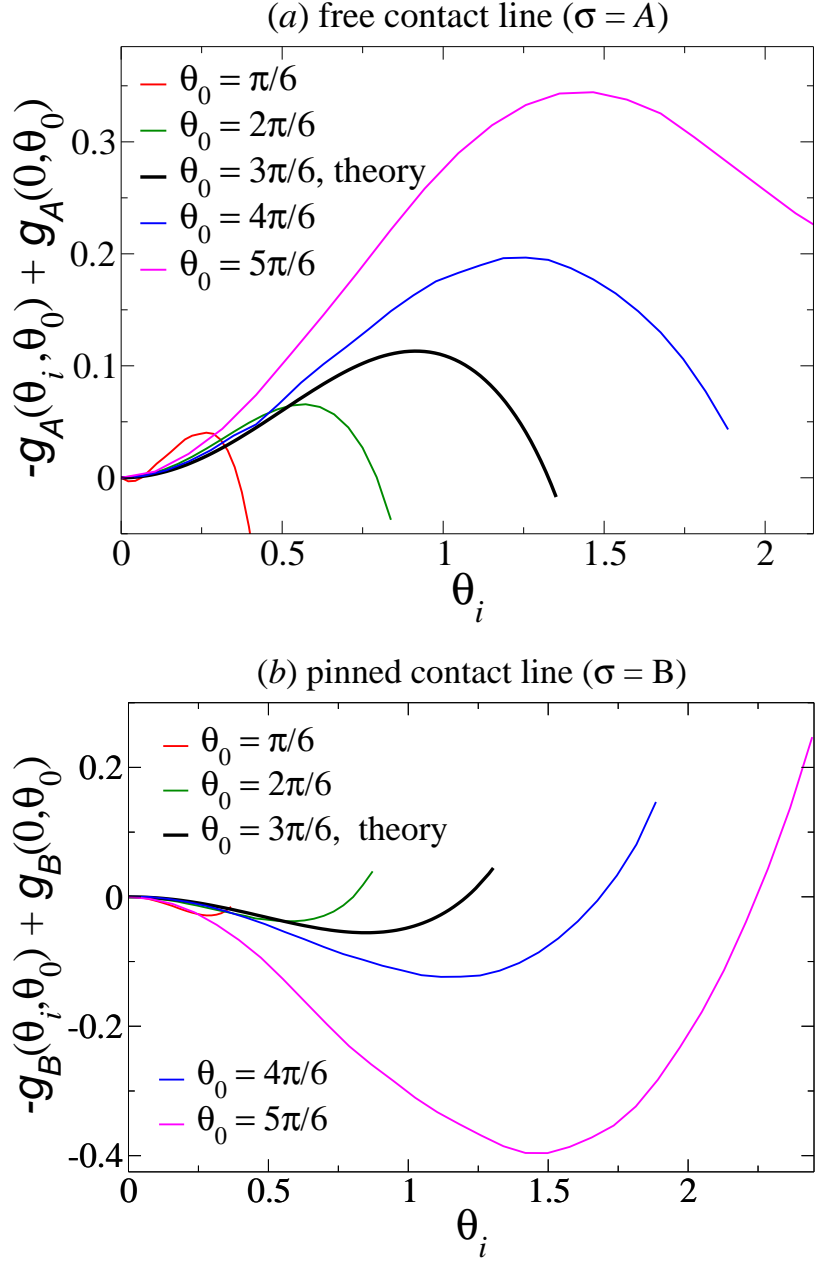

\begin{overpic}[width=0.65\textwidth]{fig09}
\end{overpic}\\
\vspace*{0.5cm}
\begin{overpic}[width=0.65\textwidth]{fig10}
\end{overpic}
\caption[]{The numerical results for the scaling functions $g_{\sigma}(\theta_i,\theta_0)$ (Eq.\ (\ref{excess_free_enND})) obtained by taking the average of the results for $f_i/(\gamma a)=\pm2$ separately for each of the several chosen values of $\theta_0$; $(a)$ free contact; $(b)$ pinned contact line. Note that for geometrical reasons $\theta_i\leq\theta_0-\delta$ with $\delta\gtrsim a/R_0$;
 \label{fig:landscapes}
}
\end{figure}

Qualitatively, the results (see Fig.\ \ref{fig:landscapes}) are the same as in the case $\theta_0=\pi/2$ (Fig.\ \ref{fig:gsigma}), i.e., for a free contact line the free energy $\Delta F^{(1)}_{A,i}(\theta_i,\theta_0)$ has two minima: at the  drop apex ($\theta_i=0$) and at the contact line ($\theta_i=\theta_0-\delta$, where $\delta$ is the closest possible angular approach of the particle to the contact line), whereas in the case of a pinned contact line $\Delta F^{(1)}_{B,i}(\theta_i,\theta_0)$ has only a single minimum at an intermediate angle $\theta_i=\theta_{min}(\theta_0)$, which strongly depends on $\theta_0$. Quantitatively, the energy barriers associated with the local free energy minima in both cases grow with $\theta_0$ approaching $\pi$. According to our analytical theory, the radial displacement of the particle $h$ is proportional to Green's function (see Eqs.\ (57) and (58) in Ref.~\cite{Guzowski2010}): $h= |f|v_{\sigma}(\Omega_i)/\gamma + O(a) = f[\ln(R_0/a)+2\pi G_{reg,\sigma}(\Omega_i,\Omega_i,\theta_0)+O(1)]/(2\pi\gamma)$ (see Eq.\ (\ref{AB-solution})). Therefore the effective elastic modulus of the droplet, defined as $k_{eff}:=df/dh=(dh/df)^{-1}$, is given by
\begin{multline}
 k_{eff}=k_{eff,\sigma}=2\pi\gamma/[\ln(R_0/a)+2\pi g_{\sigma}(\theta_i,\theta_0)+O(1)]\\
 =2\pi\gamma/\{\ln(R_0/a)+2\pi g_{\sigma}(0,\theta_0)-2\pi [-g_{\sigma}(\theta_i,\theta_0)+g_{\sigma}(0,\theta_0)]+O(1)]\}.
\label{eq:keff}
\end{multline}
Due to the good agreement of the analytical and numerical results for $\theta_0=\pi/2$ {and for droplet sizes down to $R_0/a=4$, as discussed in Ref.~\cite{Guzowski2010},} one can expect that the corrections $O(1)$ indicated in Eq.\ (\ref{eq:keff}), due to the finite size of the particle, are small. Accordingly, our numerical results in Fig. \ref{fig:landscapes} lead to the conclusion that in the case of a free contact line at the substrate ($\sigma=A$) the droplet is relatively softest if the pulling force is either directed along the $z$-axis or applied at the contact line, whereas in the case of a pinned contact line ($\sigma=B$) the droplet is softest if the force is directed along the specific intermediate polar angle $\theta_{min}(\theta_0)$.

\begin{figure}
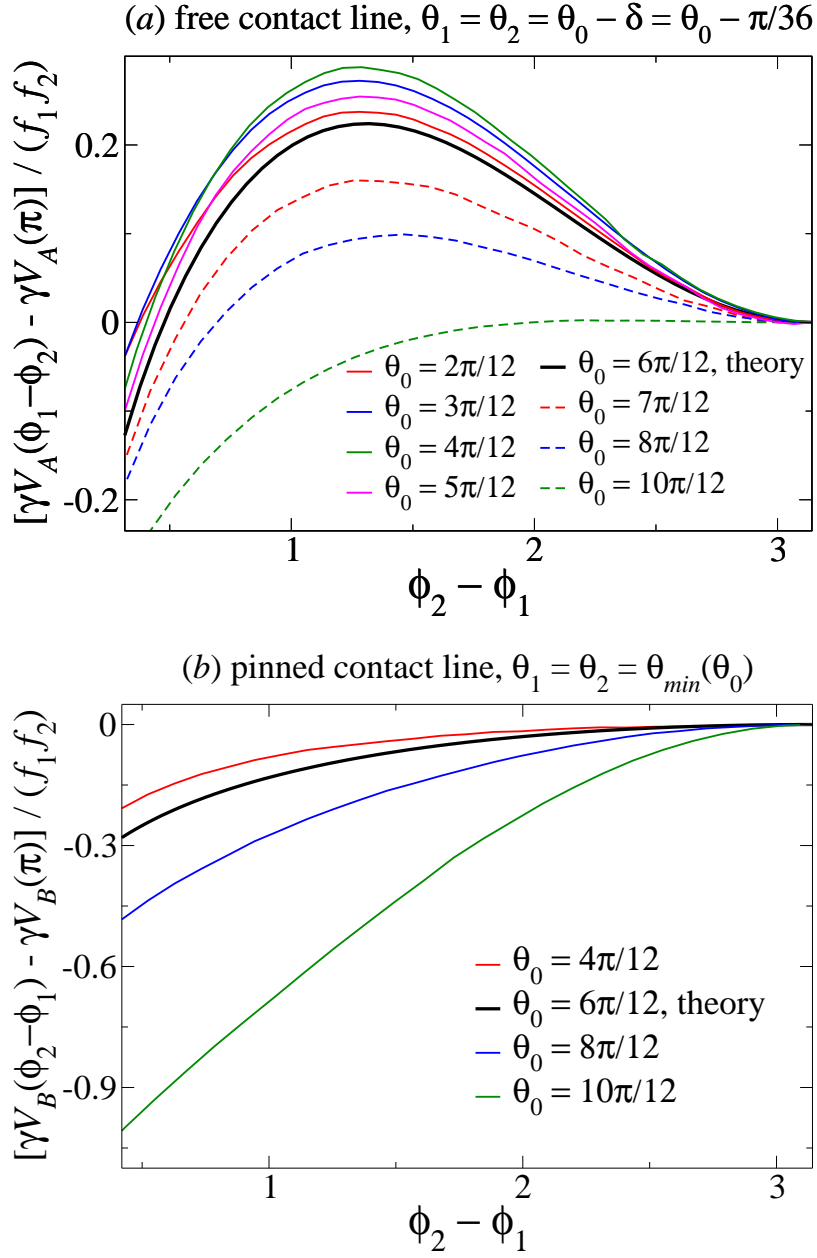

\begin{overpic}[width=0.65\textwidth]{fig11}
\end{overpic}\\
\vspace*{0.5cm}
\begin{overpic}[width=0.65\textwidth]{fig12}
\end{overpic}
\caption[]{The effective pair potentials for two particles of equal radii $a$, both placed at the polar angles $\theta_1=\theta_2$ corresponding to the minimum of the one-particle free energy landscape $\Delta F_{\sigma,i}^{(1)}$ (see Fig.\ \ref{fig:landscapes}), as a function of the relative azimuthal angle $\phi_2-\phi_1$ between the particles (see Fig.\ \ref{fig:system_sketch}) for various contact angles $\theta_0$. $(a)$ free contact line: $\theta_1=\theta_2=\theta_0-\delta$, with $\delta=\pi/36$ (Fig.\ \ref{fig:landscapes}$(a)$); $(b)$ pinned contact line: $\theta_1=\theta_2=\theta_{min}(\theta_0)$ (Fig.\ \ref{fig:landscapes}$(b)$). {The colored lines represent the numerical results for a droplet of radius $R_0/a=20$ for $\theta_0=2\pi/12$ and $\theta_0=4\pi/12$, $R_0/a=14$ for $\theta_0=3\pi/12$, and $R_0/a=12$ for the remaining contact angles. The thick black lines correspond to the analytical expression $V_{\sigma}(\phi_2-\phi_1)-V_{\sigma}(0)$ with $V_{\sigma}$ given by Eq.\ (\ref{pair-pot}).}
 \label{fig:numerical}
}
\end{figure}

In analogy to the case $\theta_0=\pi/2$, for two particles we expect that in order to reach equilibrium each particle drifts towards the polar angle $\theta_i$ close to that of the minimum of $\Delta F^{(1)}_{\sigma,i}(\theta_i,\theta_0)$. In the case of a free contact line this corresponds to a configuration in which both particles are positioned close to the drop apex or at the contact line. In the latter case we have evaluated the free energy numerically as a function of the azimuthal angle $\phi_2-\phi_1$ (see Fig.\ \ref{fig:numerical}$(a)$) 
{for $\theta_1=\theta_2=\theta_0-\delta$. We have performed the calculations for $\theta_0<\pi/2$ with $f_1=f_2>0$ and for $\theta_0>\pi/2$ with $f_1=f_2<0$, where the restriction on the sign of the force follows from the constraint that the particles cannot penetrate the substrate. Having for a given $\theta_0$ the data accessible only for one sign of the force we could not take the average; instead we had to apply a finite size correction (see Appendix A), analogous to the one proposed in Ref.~\cite{Guzowski2010}.} The data show that there are one or two minima depending on $\theta_0$. There is always a minimum for $\phi_2-\phi_1=\delta$. If the contact angle $\theta_0$ is smaller than a critical angle $\approx5\pi/6$, there is a second minimum for $\phi_2-\phi_1=\pi$. The free energy barrier between the two minima is largest for $\theta_0\approx\pi/3$ and it vanishes for $\theta_0\gtrsim5\pi/6$.
In the case of a pinned contact line (see Fig.\ \ref{fig:numerical}$(b)$) the free energy for the particles positioned at the polar angles $\theta_1=\theta_2=\theta_{min}(\theta_0)$ is, independently of $\theta_0$, always a monotonic function of $\phi_2-\phi_1$ with a minimum corresponding to the particles touching each other ($\phi_2-\phi_1=\delta$). The absence of a second minimum at $\phi_2-\phi_1=\pi$ in this case is rather surprising. For example, for $\theta_0=5\pi/6$ one has $\theta_1=\theta_2=\theta_{min}(\theta_0=5\pi/6)\simeq \pi/2$, so that for $\phi_2-\phi_1=\pi$ the angular separation $\theta_{21}$ (Fig.\ \ref{fig:system_sketch}) between the particles is almost equal to $\pi$, which, in the absence of the substrate (i.e., for a free droplet with a fixed center of mass), would indeed correspond to a local free energy minimum~\cite{comment1}. This shows that the presence of the substrate, even if the particles are far away from the contact line, can change the interactions also qualitatively.

\section{Summary and discussion}
We have obtained analytical results for the free energy of two capillary monopoles at the surface of a sessile drop (Fig.\ \ref{fig:system_sketch}). These monopoles are generated by external forces acting in radial directions. The results, which have been obtained by using the method of images for the special case of a contact angle $\theta_0=\pi/2$ but for arbitrary angular positions $\Omega_i=(\theta_i,\phi_i)$ of both monopoles $i=1,2$ (Figs.\ \ref{fig:Model_A} and \ref{fig:Model_B}), show that the preferred configuration of the particles depends sensitively on the boundary conditions at the substrate (see Fig.\ \ref{fig:gsigma}). For a free contact line the particles are attracted by the contact line (see $\theta_1=7\pi/18$ in Fig.\ \ref{fig:free_energy_vs_phi21}$(a)$ and $\theta_1=4\pi/9$ in Fig.\ \ref{fig:free_energy_vs_phi2-phi1}$(a)$). In this case two locally stable configurations are possible: the one in which the particles touch each other and the one with the particles at opposite sides of the contact line. On the contrary, for a pinned contact line the particles drift towards a characteristic polar angle $\theta_{min}\approx 49^{\circ}$ (see $\theta_1=\theta_{min}$ in Fig.\ \ref{fig:free_energy_vs_phi21}$(b)$) and there they always attract each other; in this case there is no other locally stable configuration. 
For $\theta_0=\pi/2$ the effective pair potential between a probe particle $2$ and a fixed reference particle $1$ is illustrated for various reference configurations $\Omega_1$ in Figs.\ \ref{fig:Model_A} and \ref{fig:Model_B} for a free and a pinned contact line, respectively. 

We have also obtained numerical results for an arbitrary contact angle
$\theta_0\neq\pi/2$. For a single particle $i$ the results are qualitatively
the same as for $\theta_0=\pi/2$ (compare Figs.\ \ref{fig:gsigma} and
\ref{fig:landscapes}). There is a free energy barrier between the two minima
at $\theta_i=0$ and $\theta_i=\theta_0-\delta$ for a free contact line and a
single global minimum at $\theta_i=\theta_{min}(\theta_0)$ in the case of a
pinned contact line. In the presence of two particles the corresponding
effects described above for $\theta_0=\pi/2$ are to a certain extent preserved
qualitatively. However, for $\theta_0\gtrsim5\pi/6$ and with both particles
at the free contact line, the second minimum corresponding to particles being
at opposite sides of the contact line disappears and their mutual attraction
varies purely monotonically (see Fig.\ \ref{fig:numerical}$(a)$).

In general the free energy barrier increases significantly with $\theta_0$
(see Fig.\ \ref{fig:landscapes}$(a)$). For $\theta_0=5\pi/6$ it is one order of magnitude larger then for $\theta_0=\pi/6$. The above analysis is valid for micrometer-sized or smaller droplets for which the effects of gravity can be neglected. The radial forces acting on the particles can be generated by using, e.g., optical tweezers. In this set-up the  free energy landscapes $\Delta F_{\sigma}^{(2)}$ could be determined for example by monitoring the Brownian motion of the probe particle with the reference particle pinned by an optical tweezer. Alternatively one could directly measure the tangential force on the probe particle $\nabla_a\Delta F_{\sigma}^{(2)}(\Omega_1=const,\Omega_2=\Omega)/R_0$. The first method works successfully if the particles can explore all angular configurations, which is possible only if the free energy barrier is comparable with or smaller than the thermal energy $k_BT$. For particles of equal radii $a_1=a_2=a$ on a droplet with a surface tension $\gamma\approx0.05N/m$, and subjected to radial forces $f_1=f_2=\gamma a$, the typical free energy variations across the whole droplet are of the order of $\gamma a^2/10$ (see Figs.\ \ref{fig:Model_A} and \ref{fig:Model_B}), which at a room temperature correspond to variations of the order $\Delta F_{\sigma}^{(2)}\sim 10^6 k_BT$ for $a=1\mu m$, $\sim10^4 k_BT$ for $a=100 nm$, and $\sim10^2k_BT$ for $a=10 nm$. These values are independent of the droplet radius $R_0$ as long as the latter is smaller then the capillary length. They are reduced by two orders of magnitude if the forces are reduced by one order of magnitude. In such a case particles of sizes $a\lesssim10nm$ could indeed explore the whole free energy landscape. 

In view of the possibility of experimental realizations it is instructive to discuss if the anisotropies of the interactions induced by the presence of the substrate are relevant for actual systems. For $R_0/a\simeq23$ the polar angular separation $\theta_{21}=\pi/36\simeq0.087$ corresponds to the situation that the particles almost touch each other. In the case of a pinned contact line and $\theta_0=\pi/2$ the amplitude of the free energy variations of the probe particle as a function of the local azimuthal angle $\phi_{21}$ with the reference particle positioned at $\theta_1\simeq\theta_{min}$ (see Fig.\ \ref{fig:free_energy_vs_phi21}) is given approximately by $10^{-3} \gamma a^2$. For $f_1=f_2=\gamma a$ this corresponds to $10^4 k_BT$ for $a=1\mu m$, $10^2 k_BT$ for $a=100 nm$, and $k_BT$ for $a=10 nm$. Therefore we expect that for particles of sizes $a>10nm$ the anisotropy of their interactions influences significantly the equilibrium configurations of the particles.


\begin{thebibliography}{99}
\expandafter\ifx\csname natexlab\endcsname\relax\def\natexlab#1{#1}\fi
\expandafter\ifx\csname bibnamefont\endcsname\relax
  \def\bibnamefont#1{#1}\fi
\expandafter\ifx\csname bibfnamefont\endcsname\relax
  \def\bibfnamefont#1{#1}\fi
\expandafter\ifx\csname citenamefont\endcsname\relax
  \def\citenamefont#1{#1}\fi
\expandafter\ifx\csname url\endcsname\relax
  \def\url#1{\texttt{#1}}\fi
\expandafter\ifx\csname urlprefix\endcsname\relax\def\urlprefix{URL }\fi
\providecommand{\bibinfo}[2]{#2}
\providecommand{\eprint}[2][]{\url{#2}}

\bibitem[{\citenamefont{Kraft et~al.}(2009)\citenamefont{Kraft, Groenewold, and
  Kegel}}]{Kraft2009}
\bibinfo{author}{\bibfnamefont{D.~J.} \bibnamefont{Kraft}},
  \bibinfo{author}{\bibfnamefont{J.}~\bibnamefont{Groenewold}},
  \bibnamefont{and} \bibinfo{author}{\bibfnamefont{W.~K.} \bibnamefont{Kegel}},
  \bibinfo{journal}{Soft Matter} \textbf{\bibinfo{volume}{5}},
  \bibinfo{pages}{3823} (\bibinfo{year}{2009}).

\bibitem[{\citenamefont{Aarts et~al.}(2004)\citenamefont{Aarts, Schmidt, and
  Lekkerkerker}}]{Aarts2004}
\bibinfo{author}{\bibfnamefont{D.~G. A.~L.} \bibnamefont{Aarts}},
  \bibinfo{author}{\bibfnamefont{M.}~\bibnamefont{Schmidt}}, \bibnamefont{and}
  \bibinfo{author}{\bibfnamefont{H.~N.~W.} \bibnamefont{Lekkerkerker}},
  \bibinfo{journal}{Science} \textbf{\bibinfo{volume}{304}},
  \bibinfo{pages}{847} (\bibinfo{year}{2004}).

\bibitem[{\citenamefont{Emory and Nie}(1998)}]{Emory1998}
\bibinfo{author}{\bibfnamefont{S.}~\bibnamefont{Emory}} \bibnamefont{and}
  \bibinfo{author}{\bibfnamefont{S.}~\bibnamefont{Nie}},
  \bibinfo{journal}{J.~Phys.~Chem.~B}
  \textbf{\bibinfo{volume}{102}}, \bibinfo{pages}{493} (\bibinfo{year}{1998}).

\bibitem[{\citenamefont{Aveyard et~al.}(2003)\citenamefont{Aveyard, Binks, and
  Clint}}]{Aveyard2003}
\bibinfo{author}{\bibfnamefont{R.}~\bibnamefont{Aveyard}},
  \bibinfo{author}{\bibfnamefont{B.}~\bibnamefont{Binks}}, \bibnamefont{and}
  \bibinfo{author}{\bibfnamefont{J.}~\bibnamefont{Clint}},
  \bibinfo{journal}{Adv. Colloid Interface Sci.}
  \textbf{\bibinfo{volume}{100}}, \bibinfo{pages}{503} (\bibinfo{year}{2003}).

\bibitem[{\citenamefont{Zahn and Maret}(2000)}]{Zahn2000}
\bibinfo{author}{\bibfnamefont{K.}~\bibnamefont{Zahn}} \bibnamefont{and}
  \bibinfo{author}{\bibfnamefont{G.}~\bibnamefont{Maret}},
  \bibinfo{journal}{Phys.\ Rev.\ Lett.} \textbf{\bibinfo{volume}{85}},
  \bibinfo{pages}{3656} (\bibinfo{year}{2000}).

\bibitem[{\citenamefont{Kosterlitz and Thouless}(1973)}]{Kosterlitz1973}
\bibinfo{author}{\bibfnamefont{J.~M.} \bibnamefont{Kosterlitz}}
  \bibnamefont{and} \bibinfo{author}{\bibfnamefont{D.~J.}
  \bibnamefont{Thouless}}, \bibinfo{journal}{J.\ Phys.\ C:\ Solid State Phys.}
  \textbf{\bibinfo{volume}{6}}, \bibinfo{pages}{1181} (\bibinfo{year}{1973}).

\bibitem[{\citenamefont{Hurd}(1985)}]{HURD1985}
\bibinfo{author}{\bibfnamefont{A.~J.} \bibnamefont{Hurd}},
  \bibinfo{journal}{J.\ Phys.\ A: Math.\ Gen.} \textbf{\bibinfo{volume}{18}},
  \bibinfo{pages}{1055} (\bibinfo{year}{1985}).

\bibitem[{\citenamefont{Pieranski}(1980)}]{Pieranski1980}
\bibinfo{author}{\bibfnamefont{P.}~\bibnamefont{Pieranski}},
  \bibinfo{journal}{Phys.\ Rev.\ Lett.} \textbf{\bibinfo{volume}{45}},
  \bibinfo{pages}{569} (\bibinfo{year}{1980}).

\bibitem[{\citenamefont{Ruiz-Garcia et~al.}(1998)\citenamefont{Ruiz-Garcia,
  Gamez-Corrales, and Ivlev}}]{Ruiz-Garcia1998}
\bibinfo{author}{\bibfnamefont{J.}~\bibnamefont{Ruiz-Garcia}},
  \bibinfo{author}{\bibfnamefont{R.}~\bibnamefont{Gamez-Corrales}},
  \bibnamefont{and} \bibinfo{author}{\bibfnamefont{B.~I.} \bibnamefont{Ivlev}},
  \bibinfo{journal}{Phys.\ Rev.\ E} \textbf{\bibinfo{volume}{58}},
  \bibinfo{pages}{660} (\bibinfo{year}{1998}).

\bibitem[{\citenamefont{Ghezzi et~al.}(2001)\citenamefont{Ghezzi, Earnshaw,
  Finnis, and McCluney}}]{Ghezzi2001}
\bibinfo{author}{\bibfnamefont{F.}~\bibnamefont{Ghezzi}},
  \bibinfo{author}{\bibfnamefont{J.~C.} \bibnamefont{Earnshaw}},
  \bibinfo{author}{\bibfnamefont{M.}~\bibnamefont{Finnis}}, \bibnamefont{and}
  \bibinfo{author}{\bibfnamefont{M.}~\bibnamefont{McCluney}},
  \bibinfo{journal}{J.\ Colloid Interface Sci.} \textbf{\bibinfo{volume}{238}},
  \bibinfo{pages}{433} (\bibinfo{year}{2001}).

\bibitem[{\citenamefont{Nikolaides et~al.}(2002)\citenamefont{Nikolaides,
  Bausch, Hsu, Dinsmore, Brenner, Weitz, and Gay}}]{Nikolaides2002}
\bibinfo{author}{\bibfnamefont{M.~G.} \bibnamefont{Nikolaides}},
  \bibinfo{author}{\bibfnamefont{A.~R.} \bibnamefont{Bausch}},
  \bibinfo{author}{\bibfnamefont{M.~F.} \bibnamefont{Hsu}},
  \bibinfo{author}{\bibfnamefont{A.~D.} \bibnamefont{Dinsmore}},
  \bibinfo{author}{\bibfnamefont{M.~P.} \bibnamefont{Brenner}},
  \bibinfo{author}{\bibfnamefont{D.~A.} \bibnamefont{Weitz}}, \bibnamefont{and}
  \bibinfo{author}{\bibfnamefont{C.}~\bibnamefont{Gay}},
  \bibinfo{journal}{Nature} \textbf{\bibinfo{volume}{420}},
  \bibinfo{pages}{299} (\bibinfo{year}{2002}).

\bibitem[{\citenamefont{Oettel et~al.}(2005)\citenamefont{Oettel,
  Dom\'{\i}nguez, and Dietrich}}]{Oettel2005}
\bibinfo{author}{\bibfnamefont{M.}~\bibnamefont{Oettel}},
  \bibinfo{author}{\bibfnamefont{A.}~\bibnamefont{Dom\'{\i}nguez}},
  \bibnamefont{and} \bibinfo{author}{\bibfnamefont{S.}~\bibnamefont{Dietrich}},
  \bibinfo{journal}{Phys.\ Rev.\ E} \textbf{\bibinfo{volume}{71}},
  \bibinfo{pages}{051401} (\bibinfo{year}{2005}).

\bibitem[{\citenamefont{Dom\'{\i}nguez
  et~al.}(2007)\citenamefont{Dom\'{\i}nguez, Oettel, and
  Dietrich}}]{Dominguez2007a}
\bibinfo{author}{\bibfnamefont{A.}~\bibnamefont{Dom\'{\i}nguez}},
  \bibinfo{author}{\bibfnamefont{M.}~\bibnamefont{Oettel}}, \bibnamefont{and}
  \bibinfo{author}{\bibfnamefont{S.}~\bibnamefont{Dietrich}},
  \bibinfo{journal}{J.\ Chem.\ Phys.} \textbf{\bibinfo{volume}{127}},
  \bibinfo{pages}{204706} (\bibinfo{year}{2007}).

\bibitem[{\citenamefont{Guzowski et~al.}()\citenamefont{Guzowski, Tasinkevych,
  and Dietrich}}]{Guzowski2010}
\bibinfo{author}{\bibfnamefont{J.}~\bibnamefont{Guzowski}},
  \bibinfo{author}{\bibfnamefont{M.}~\bibnamefont{Tasinkevych}},
  \bibnamefont{and} \bibinfo{author}{\bibfnamefont{S.}~\bibnamefont{Dietrich}},
  \bibinfo{journal}{Eur.\ Phys.\ J.\ E} \textbf{\bibinfo{volume}{33}},
  \bibinfo{pages}{219} (\bibinfo{year}{2010}).

\bibitem[{\citenamefont{Kralchevsky and Nagayama}(2001)}]{Kralchevsky_book}
\bibinfo{author}{\bibfnamefont{P.~A.} \bibnamefont{Kralchevsky}}
  \bibnamefont{and} \bibinfo{author}{\bibfnamefont{K.}~\bibnamefont{Nagayama}},
  \emph{\bibinfo{title}{Particles at Fluid Interfaces}}
  (\bibinfo{publisher}{Elsevier}, \bibinfo{address}{Amsterdam},
  \bibinfo{year}{2001}).

\bibitem[{\citenamefont{Dom\'{\i}nguez
  et~al.}(2008)\citenamefont{Dom\'{\i}nguez, Oettel, and
  Dietrich}}]{Dominguez2008a}
\bibinfo{author}{\bibfnamefont{A.}~\bibnamefont{Dom\'{\i}nguez}},
  \bibinfo{author}{\bibfnamefont{M.}~\bibnamefont{Oettel}}, \bibnamefont{and}
  \bibinfo{author}{\bibfnamefont{S.}~\bibnamefont{Dietrich}},
  \bibinfo{journal}{J.\ Chem.\ Phys.} \textbf{\bibinfo{volume}{128}},
  \bibinfo{pages}{114904} (\bibinfo{year}{2008}).

\bibitem[{\citenamefont{Morse and Witten}(1993)}]{Morse1993}
\bibinfo{author}{\bibfnamefont{D.~C.} \bibnamefont{Morse}} \bibnamefont{and}
  \bibinfo{author}{\bibfnamefont{T.~A.} \bibnamefont{Witten}},
  \bibinfo{journal}{Europhys.\ Lett.} \textbf{\bibinfo{volume}{22}},
  \bibinfo{pages}{549} (\bibinfo{year}{1993}).

\bibitem[{\citenamefont{Brakke}(1992)}]{Brakke}
\bibinfo{author}{\bibfnamefont{K.}~\bibnamefont{Brakke}},
  \bibinfo{journal}{Experimental Mathematics} \textbf{\bibinfo{volume}{1}},
  \bibinfo{pages}{141�165} (\bibinfo{year}{1992}).


\bibitem[{com()}]{comment1}
\bibinfo{note}{Due to symmetry arguments the free energy of two particles on a full spherical droplet is twice the free energy of a single particle on a sessile droplet with a free contact line and $\theta_0=\pi/2$, which exhibits a deep second minimum (see Eq.\ (\ref{excess_free_enND}) and Fig.\ \ref{fig:gsigma}).}

\end{thebibliography}

\appendix
\section{Finite size correction to the numerically calculated free energy}

{In this appendix we calculate a finite size correction to the numerically calculated free energy. In the reference configuration ($f_1=f_2=0$) the immersed part $\delta V_{ref,i}$ of particle $i$ (being the intersection of the domain occupied by the particle with the spherical cap, representing the reference droplet of volume $V_l=(4\pi f_0(\theta_0) R_0^3/3)(1+O(a/R_0)^3)$ with $f_0(\theta)=(2+\cos\theta)(1-\cos\theta)^2/4$) has, independently of $\theta_i$ and for the contact angle at the particle equal to $\pi/2$, the volume $\delta V_{ref,i}=(2\pi a^3/3)(1+O(a/R_0))$. The position of this cavity in the liquid depends on $\Omega_i$. Therefore the position $x_{CM,ref}$ of the center of mass of liquid depends on $\Omega_1$ and $\Omega_2$. In the case $\theta_2=\theta_1$ it equals (here the direction of the $x$-axis is taken such that $\phi_2=\Delta\phi/2$ and $\phi_1=-\Delta\phi/2$, where $\Delta\phi\equiv\phi_2-\phi_1$)
\begin{multline}
x_{CM,ref}(\theta_1,\phi_2-\phi_1)=\dfrac{\int_{V_l}\!dV\, x}{V_l}=-\dfrac{1}{V_l}\int_{\delta V_{ref,1}+\delta V_{ref,2}}\!dV\, x \\
\approx -\dfrac{2 \delta V_{ref,1}}{V_l}R_0\sin\theta_1\cos[(\phi_2-\phi_1)/2]\approx -\dfrac{a^3}{R_0^2f_0(\theta_0)}\sin\theta_1\cos[(\phi_2-\phi_1)/2].
\label{x_cm}
\end{multline}
This gives rise to the free energy contribution $\delta F$ which can be understood as the total work done by the force 
$f_{CM}(\theta_1,\phi_2-\phi_1)=-2f_1\sin\theta_1\cos[(\phi_2-\phi_1)/2]$ applied to the center of mass in order to counterbalance the lateral component of the forces $f_1$ and $f_2$ (below we assume $f_1=f_2$) upon displacing the center of mass by $-x_{CM,ref}(\theta_1,\phi_2-\phi_1)$:
\begin{multline}
	\delta F = \int_{\pi}^{\phi_2-\phi_1}\!d\phi\, f_{CM}(\phi)[-d x_{CM,ref}(\theta_1,\phi)/d\phi]\\ 
	=\dfrac{f_1a^3}{R_0^2f_0(\theta_0)}\sin^2\theta_1\int_{\pi}^{\phi_2-\phi_1}\!d\phi\,\sin(\phi/2)\cos(\phi/2)\\
	=-\dfrac{f_1a^3}{R_0^2f_0(\theta_0)}\sin^2\theta_1\cos^2[(\phi_2-\phi_1)/2]
\end{multline}
The results presented in Fig.\ \ref{fig:numerical}$(a)$ have been obtained by subtracting this correction $\delta F$ from the numerically calculated free energies.}

\end{document}